\documentstyle[12pt]{article}
\begin{document}
\begin{flushright}
IHEP 94-63\\
31 May 1994\\
\end{flushright}
\begin{center}
{\bf \large Scaling behaviour of leptonic decay constants\\ for
heavy quarkonia and heavy mesons}\\
\vspace*{0.5cm}
V.V.Kiselev\\
Institute for High Energy Physics,\\
Protvino, Moscow Region, 142284, Russia.\\
\end{center}

\begin{abstract}

In the framework of QCD sum rules one uses a scheme, allowing one to apply
the conditions of both nonrelativistic heavy quark motion inside mesons and
independence of nonsplitting nS-state density on the heavy quark flavours.
In the leading order an analitic expression is derived for leptonic
constants of both heavy quarkonia and heavy mesons with a single heavy
quark. The expression allows one explicitly to determine scaling properties
of the constants.

\end{abstract}

\section*{Introduction}

By definition, heavy quarks have the $m_Q$ mass values that are much greater
than the confinement energy $\Lambda$. Therefore, in some cases,
consideration of hadrons with the heavy quarks allows one to use expansions
of some quantities over the small parameter of the $\Lambda/m_Q$ ratio.
If the heavy quark virtualities inside a hadron are not large, then one
allows the kinematical expansion
\begin{eqnarray}
p_Q^\mu & = & m_Q\cdot v^\mu + k^\mu\;, \nonumber \\
v\cdot k & \sim & 0\;, \label{1n}\\
|k^2| & \ll & m_Q^2\;, \nonumber
\end{eqnarray}
where $v$ is the quark 4-velocity, $p_Q$ is the quark momentum, $k$
characterizes the heavy quark virtuality. In the kinematics of eq.(\ref{1n})
in QCD, the heavy quark action, expanded over the small parameter,
leads to Effective Heavy Quark Theory (EHQT) \cite{1}, so, in the leading
approximation the theory possesses the symmetry with respect to the
substitution of a heavy quark, moving with the velocity $\vec{v}$,
by any other heavy quark, moving with the same velocity $\vec{v}$
and having an arbitrary orientation of its spin.

In the system, where $v=(1,\vec{0})$, the heavy quark hamiltonian in
an external field has the form
\begin{equation}
H = m_Q + V(r) + \frac{\vec{k}^2}{2m_Q} + g \frac{\vec{\sigma}\cdot \vec{B}}
{2m_Q} + O(1/m_Q^2)\;, \label{2n}
\end{equation}
where $V$ is the potential, $\vec{B}$ is the chromomagnetic field.

In heavy mesons $(Q\bar q)$ with a single heavy quark, one has
\begin{eqnarray}
\frac{<\vec{k}^2>}{m_Q} & \sim & O(1/m_Q)\;, \label{3n}\\
g \frac{<\vec{\sigma}\cdot \vec{B}>}{m_Q} & \sim & O(1/m_Q)\;, \nonumber
\end{eqnarray}
and the distance $r$ is determined by the light quark motion around the static
source of the gluon field. Hence, in the leading approximation of EHQT,
the heavy meson wave function is universal and independent of the flavour
of the heavy quark inside the meson. This feature leads to both
the scaling law for the leptonic constants of the heavy mesons
\begin{equation}
f^2\cdot M = const.\;, \label{4n}
\end{equation}
and the universality of form factors for semileptonic transitions between
the hadrons, containing a single heavy quark (for example,
$B \to D^{(*)} l \nu$) \cite{1}.

In the case of heavy quarkonium $(Q\bar Q')$, the chromomagnrtic field arises
only at nonzero velocity of the source, so that
\begin{equation}
\vec{B} \sim O(\vec{v}) \sim O(1/m_Q)\;, \label{5n}
\end{equation}
and, hence, spin-dependent splittings of the quarkonium levels arise only at
the second order over $1/m_Q$, so that in what follows, we neglect the
spin-dependent splittings in the heavy quarkonium.

As for the kinetic energy of the heavy quark motion inside the quarkonium
$(Q\bar Q')$
\begin{equation}
T = \frac{\vec{k}^2}{2m_Q} + \frac{\vec{k}'^2}{2m_Q}\;, \label{6n}
\end{equation}
it determines essentially both the quark binding energy $E$ and the wave
function $\Psi_E(\vec{r})$, as one can know this fact from an experience of
working with nonrelativistic potential models of the quarkonia.
Moreover, the distance between the heavy quarks inside the quarkonium
depends on the quark masses. Thus, generally speaking, the leading
approximation of EHQT may not be applied to the heavy quarkonium $(Q\bar Q')$,
whose wave function depends on the quarkonium content.

However, as it has been shown in ref.\cite{2}, in the region of
average distances between the heavy quarks inside the $(c\bar c)$
charmonium and the $(b\bar b)$ bottomonium
\begin{equation}
0.1\;fm < r < 1\;fm\;, \label{7n}
\end{equation}
and with accuracy up to an additive shift, the QCD-mo\-ti\-va\-ted
fla\-vour-\-in\-de\-pen\-dent heavy quark potentials, behaving as the Coulomb
interaction at small distances and having linearly rising confining part
at large distances (Cornell model \cite{3}, Richardson potential \cite{4},
Buchm\" uller-Tye model \cite{5}), allow the parameterizations in the forms
of logarithmic \cite{6} and power \cite{7} laws, possessing simple scaling
properties
\begin{eqnarray}
V_L(r) & = & c_L + d_L\ln{\Lambda r}\;, \label{8n}\\
V_M(r) & = & -c_M + d_M (\Lambda r)^k\;.\label{9n}
\end{eqnarray}
By the virial theorem for average values of the kinetic energies in
potentials (\ref{8n}), (\ref{9n}), one can get
\begin{eqnarray}
<T_L> & = & d_L/2 = const.\;, \label{99n}\\
<T_M> & = & \frac{k}{k+2}(c_M+E)\;, \label{10n}
\end{eqnarray}
respectively, so that at $|E|\ll c_M$, $k \ll 1$ and with the accuracy by
the small binding energy of the quarks inside the quarkonium, one
concludes
\begin{equation}
<T_M> \simeq const. \label{11n}
\end{equation}
In accordance with the Feynman-Hellmann theorem
\begin{equation}
\frac{dE}{d\mu} = -\;\frac{<T>}{\mu}\;, \label{12n}
\end{equation}
where $\mu$ is the reduced mass of the heavy quarks $(Q\bar Q')$, and
by condition (\ref{99n}), for the difference of the energies of two levels,
one gets
\begin{equation}
E(\bar n, \mu) - E(n,\mu) = E(\bar n, \mu') - E(n, \mu')\;, \label{13n}
\end{equation}
i.e. the level density of the $(Q\bar Q')$ system does not depend on the
heavy quark flavours
\begin{equation}
\frac{dn}{dM_n} = const.\;,\label{14n}
\end{equation}
that is rather accurately confirmed  empirically \cite{8} (see table \ref{t1}).

\begin{table}[t]
\caption{The Mass Difference (in MeV) for the Lightest Vector States with The
Prescribed Valence Quark Contents.}
\label{t1}
\begin{center}
\begin{tabular}{||c|c|c|c||}
\hline
state & $\Upsilon$ & $\psi$ & $\phi$\\
\hline
$\Delta M$ & 563 & 588 & 660\\
\hline
\end{tabular}
\end{center}
\end{table}

In the framework of the QCD sum rules \cite{9} the use of
\begin{description}
\item[1)]
the small parameter , $\Lambda/m_Q \ll 1$,
\item[2)]
the nonrelativistic motion of the heavy quarks, $v\to 0$,
\item[3)]
the universality of the quarkonium state density (\ref{14n}),
\end{description}
allows one, in the leading order,
\begin{description}
\item[1)]
to neglect power $\Lambda/m_Q$ corrections from quark-gluon condensates,
\item[2)]
to take into the account Coulomb-like interactions over $\alpha_S/v$,
\item[3)]
to derive the scaling relation for the leptonic decay constants of heavy S-wave
quarkonium (see ref.\cite{10})
\begin{equation}
\frac{f^2}{M} =  const.\;,\label{15n}
\end{equation}
in the regime, when $|m_Q-m_{Q'}|$ is restricted at $m_{Q,Q'} \gg \Lambda$.
Expression (\ref{15n}) is in a good agreement with the experimental values of
$f_\Upsilon$, $f_\psi$ and $f_\phi$.
\end{description}

In the present paper we generalize the analisys, made in ref.\cite{10},
for the regime of $m_Q= x m_{Q'} \gg \Lambda$ and derive the following
expression, determining the scaling properties of the leptonic constants
for the S-wave quarkonia
\begin{equation}
\frac{f^2}{M}\cdot \biggl(\frac{M}{4\mu}\biggr)^2 =  const.\;,\label{16n}
\end{equation}
where $\mu$ is the reduced mass of the quarks.

Expression (\ref{16n})
\begin{description}
\item[1)]
is reduced to eq.(\ref{15n}) at $x=1$, i.e. at $4\mu/M\simeq 1$,
\item[2)]
agrees with the scaling law for the leptonic constants of heavy
$(Q\bar q)$ mesons in the regime $M\to \infty$, $\mu=const.$, and
\item[3)]
allows one to predict the $f_{B_C}$ value for the heavy quarkonium $B_c$
with the open charm and beauty (the $B_c$ search is processed at LEP and FNAL).
\end{description}

In Section 1 the QCD sum rule scheme is considered. It allows one explicitly
to use condition (\ref{14n}) and the $dn/dM_n$ quantity as the
phenomenological parameter, and to avoid unphysical dependence of results on
the external parameters such as the number of the spectral density moment
or the Borel transformation parameter. Expression (\ref{16n}) is derived.

In Section 2 one analyses the scaling relation (\ref{16n}), and in Conclusion
the obtained results are discussed.

\section{Quarkonium sum rules}

Let us consider the two-point correlator functions of the quark currents
\begin{eqnarray}
\Pi_{\mu\nu} (q^2) & = & i \int d^4x e^{iqx} <0|T J_{\mu}(x)
J^{\dagger}_{\nu}(0)|0>\;,
\label{1} \\
\Pi_P (q^2) & = & i \int d^4x e^{iqx} <0|T J_5(x) J^{\dagger}_5(0)|0>\;,
\end{eqnarray}
where
\begin{eqnarray}
J_{\mu}(x) & = & \bar Q_1(x) \gamma_{\mu} Q_2(x)\;,\\
J_5(x) & = & \bar Q_1(x) \gamma_5 Q_2(x)\;,\\
\end{eqnarray}
$Q_i$ is spinor field of the heavy quark with $i = c, b$.

Further, write down
\begin{equation}
\Pi_{\mu\nu} = \biggl(-g_{\mu\nu}+\frac{q_{\mu} q_{\nu}}{q^2}\biggr) \Pi_V(q^2)
+ \frac{q_{\mu} q_{\nu}}{q^2} \Pi_S(q^2)\;,
\end{equation}
where $\Pi_V$ and $\Pi_S$ are the vector and scalar correlator functions,
respectively. In the following we will consider the vector and pseudoscalar
correlators: $\Pi_V(q^2)$ and $\Pi_P(q^2)$.

Define the leptonic constants $f_V$ and $f_P$
\begin{eqnarray}
<0|J_{\mu}(x) |V(\lambda)> & = & i \epsilon^{(\lambda)}_{\mu}\;f_V
M_V\;e^{ikx}\;,\\
<0|J_{5\mu}(x)|P> & = & i k_{\mu}\;f_P e^{ikx}\;,
\end{eqnarray}
where
\begin{equation}
J_{5\mu}(x)  =  \bar Q_1(x) \gamma_5 \gamma_{\mu} Q_2(x)\;,
\end{equation}
so that
\begin{equation}
<0|J_{5}(x)|P>  =  i\;\frac{f_P M_P^2}{m_1+m_2}\;e^{ikx}\;, \label{9}
\end{equation}
where $|V>$ and  $|P>$ are the state vectors of the $1^-$ and $0^-$
quarkonia, and $\lambda$ is the vector quarkonium polarization, $k$
is 4-momentum of the meson, $k_{P,V}^2 = M_{P,V}^2$.

Considering the charmonium ($\psi$, $\psi '$ ...) and bottomonium ($\Upsilon$,
$\Upsilon '$, $\Upsilon ''$ ...), one can easily show that the relation
between the width of the
leptonic decay $V \to e^+ e^-$  and $f_V$ has the form
\begin{equation}
\Gamma (V \to e^+ e^-) = \frac{4 \pi}{9}\;e_i^2
\alpha_{em}^2\;\frac{f_V^2}{M_v}\;,
\end{equation}
where $e_i$ is the electric charge of the quark $i$.

In the region of the narrow nonoverlapping resonances, it follows from
Eqs.(\ref{1}) - (\ref{9}) that
\begin{eqnarray}
\frac{1}{\pi} \Im m \Pi_V^{(res)} (q^2) & = & \sum_n f_{Vn}^2 M_{Vn}^2
\delta(q^2-M_{Vn}^2)\;,
\label{11} \\
\frac{1}{\pi} \Im m \Pi_P^{(res)} (q^2) & = & \sum_n f_{Pn}^2
M_{Pn}^4\;\frac{1}{(m_1+m_2)^2} \delta(q^2-M_{Pn}^2)\;,
\end{eqnarray}
Thus, for the observed spectral function one has
\begin{equation}
\frac{1}{\pi} \Im m \Pi_{V,P}^{(had)} (q^2)  = \frac{1}{\pi} \Im m
\Pi_{V,P}^{(res)} (q^2)+ \rho_{V,P}(q^2 \mu_{V,P}^2)\;,
\label{13}
\end{equation}
where $\rho (q^2,\;\mu^2)$ is the continuum contribution, which is
not equal to zero at $q^2 > \mu^2$.

Moreover, the operator product expansion gives
\begin{equation}
\Pi^{(QCD)} (q^2)  = \Pi^{(pert)} (q^2)+ C_G(q^2) <\frac{\alpha_S}{\pi} G^2> +
C_i(q^2)<m_i \bar Q_i Q_i>+ \dots\;,
\label{14}
\end{equation}
where the perturbative contribution $\Pi^{(pert)}(q^2)$ is labled, and
the nonperturbative one is expressed in the form of the quark-gluon condensate
sum with the Wilson's coefficients, which may be calculated in the QCD
perturbative theory.

In Eq.(\ref{14}) we were restricted by the contribution of the vacuum
expectation values for the operators with dimension $d =4$.
For $C^{(P)}_G (q^2)$ one has, for instance, \cite{9}
\begin{equation}
C_G^{(P)} = \frac{1}{192 m_1 m_2}\;\frac{q^2}{\bar q^2}\;
\biggl(\frac{3(3v^2+1)(1-v^2)^2}
{2v^5} \ln \frac{1+v}{1-v} - \frac{9v^4+4v^2+3}{v^4}\biggr)\;, \label{15}
\end{equation}
where
\begin{equation}
\bar q^2 = q^2 - (m_1-m_2)^2\;,\;\;\;\;v^2 = 1-\frac{4m_1 m_2}{\bar q^2}\;.
\label{16}
\end{equation}
The analogous formulae for other Wilson's coefficients can be found in
Ref.\cite{9}. In the following it will be clear that the explicit form
of the coefficients has no significant meaning for the present consideration.

In the leading order of the QCD perturbation theory it was found for
the imaginary part of the correlator that \cite{9}
\begin{eqnarray}
\Im m \Pi_V^{(pert)} (q^2) & = & \frac{\tilde s}{8 \pi s^2} (3 \bar s s - \bar
s^2 + 6m_1 m_2 s - 2 m_2^2 s) \theta(s-(m_1+m_2)^2)\;,\\
\Im m \Pi_P^{(pert)} (q^2) & = & \frac{3 \tilde s}{8 \pi s^2} (s - (m_1-m_2)^2)
\theta(s-(m_1+m_2)^2)\;,
\end{eqnarray}
where $\bar s = s-m_1^2+m_2^2$, $ \tilde s^2 = \bar s^2 -4 m_2^2 s$.

The one-loop contribution into $\Im m \Pi(q^2)$ can be included into the
consideration (see, for example, Ref.\cite{9}). However, we note that the
more essential correction is that of summing a set over the powers of
$(\alpha_s/v)$, where $v$ is defined in Eq.(\ref{16}) and is a relative quark
velocity, and $\alpha_S$ is the QCD interaction constant. In Ref.\cite{9}
it has been shown that the account of the coulomb-like gluonic
interaction between the quarks leads to the factor
\begin{equation}
F(v) = \frac{4 \pi}{3}\;\frac{\alpha_S}{v}\; \frac{1}{1-\exp (-\frac{4 \pi
\alpha_S}{3 v})}\;,
\end{equation}
so that the expansion of the $F(v)$ over $\alpha_S/v \ll 1$ restores,
precisely,
the one-loop $O(\frac{\alpha_S}{v})$ correction
\begin{equation}
F(v) \approx 1 - \frac{2 \pi}{3}\;\frac{\alpha_s}{v}\; \dots \label{20}
\end{equation}
In accordance with the dispersion relation one has the QCD sum rules,
which state that, in average, it is true that, at least, at $q^2 < 0$
\begin{equation}
\frac{1}{\pi}\;\int\frac{\Im m \Pi^{(had)}(s)}{s-q^2} ds = \Pi^{(QCD)}(q^2)\;,
 \label{21}
\end{equation}
where the necessary subtractions are omitted. $\Im m \Pi^{(had)}(q^2)$ and
 $\Pi^{(QCD)}(q^2)$ are defined by Eqs.(\ref{11}) - (\ref{13}) and
Eqs.(\ref{14}) - (\ref{20}), respectively.
Eq.(\ref{21}) is the base to develop the sum rule approach in the forms
of the correlator function moments and of the Borel transform analysis
(see Ref.\cite{9}). The truncation of the set in the right hand side of
Eq.(\ref{21}) leads to the mentioned unphysical dependence of the $f_{P,V}$
values on the external parameter of the sum rule scheme.

Further, let us use the conditions, simplifying the consideration due to
the heavy quarkonium.

\subsection{Nonperturbative Contribution}

We assume that, in the limit of the very heavy quark mass, the power
corrections of the nonperturbative contribution are small. From Eq.(\ref{15})
one can see that, for example,
\begin{equation}
C_G^{(P)}(q^2) \approx O(\frac{1}{m_1 m_2})\;,\;\; \Lambda/m_{1,2}\ll 1\;,
\end{equation}
where $v$ is fixed,  $q^2 \sim (m_1 + m_2)^2$,
when $\Im m \Pi^{(pert)}(q^2) \sim (m_1+m_2)^2$.
It is evident that, due to the purely dimensional consideration, one can
believe that the Wilson's coefficients tend to zero as
$1/m_{1,2}^2$.

Thus, the limit of the very large heavy quark mass implies that one can neglect
the quark-gluon condensate contribution.

\subsection{Nonrelativistic Quark Motion}

The nonrelativistic quark motion implies that, in the resonant region, one has,
in accordance with Eq.(\ref{16}), that
\begin{equation}
v \to 0\;.
\end{equation}
So, one can easily find that in the leading order
\begin{equation}
\Im m \Pi_P^{(pert)}(s) \approx  \Im m \Pi_V^{(pert)}(s) \to \frac {3 v}
{8 \pi^2} s\; \biggl(\frac{4\mu}{M}\biggr)^2\;,
\end{equation}
so that with account of the coulomb factor
\begin{equation}
F(v) \simeq \frac{4 \pi}{3}\; \frac{\alpha_S}{v}\;,
\end{equation}
one obtaines
\begin{equation}
\Im m \Pi_{P,V}^{(pert)}(s) \simeq \frac{\alpha_S}{2} s\;
\biggl(\frac{4\mu}{M}\biggr)^2\;. \label{27}
\end{equation}

\subsection{"Smooth Average Value" Scheme of the Sum Rules}

As for the hadronic part of the correlator, one can write down for the narrow
vector resonance contribution
\begin{eqnarray}
\Pi_V^{(res)}(q^2) & = & \int \frac{ds}{s-q^2}\;\sum_n f^2_{Vn} M^2_{Vn}
\delta(s-M_{Vn}^2)\;,
\label{28} \\
\Pi_P^{(res)}(q^2) & = & \int \frac{ds}{s-q^2}\;\sum_n f^2_{Pn}
\frac{M^4_{Pn}}{(m_1+m_2)^2} \delta(s-M_{Pn}^2)\;,\label{29}
\end{eqnarray}
The integrals in Eqs.(\ref{28})-(\ref{29}) are simply calculated, and
this procedure is generally used.

In the presented scheme, let us introduce the function of the state number
$n(s)$, so that
\begin{equation}
n(m_k^2) = k\;.
\end{equation}
This definition seems to be reasonable in the resonant region.
Then one has, for example, that
\begin{equation}
\frac{1}{\pi}\; \Im m \Pi_V^{(res)}(s) = s f^2_{Vn(s)}\; \frac{d}{ds} \sum_k
\theta(s-M^2_{Vk})\;.
\end{equation}
Further, it is evident that
\begin{equation}
\frac{d}{ds} \sum_k \theta(s-M_k^2) = \frac{dn(s)}{ds}\;\frac{d}{dn} \sum_k
\theta(n-k)\;,
\end{equation}
and Eq.(\ref{28}) may be rewritten as
\begin{equation}
\Pi_V^{(res)}(q^2) = \int \frac{ds}{s-q^2}\; s f^2_{Vn(s)}\;\frac{dn(s)}{ds}\;
\frac{d}{dn} \sum_k \theta(n-k)\;.
\end{equation}
The "smooth average value" scheme means that
\begin{equation}
\Pi_V^{(res)}(q^2) = <\frac{d}{dn} \sum_k \theta(n-k)>\; \int \frac{ds}{s-q^2}
s f^2_{Vn(s)} \frac{dn(s)}{ds}\;.
\end{equation}
It is evident that, in average, the first derivative of the step-like function
in the resonant region is equal to
\begin{equation}
<\frac{d}{dn} \sum_k \theta(n-k)> \simeq 1\;.
\end{equation}
Thus, in the scheme one has
\begin{eqnarray}
<\Pi_V^{(res)}(q^2)> & \approx & \int \frac{ds}{s-q^2} s f^2_{Vn(s)}\;
\frac{dn(s)}{ds}\;,
\label{35} \\
<\Pi_P^{(res)}(q^2)> & \approx & \int \frac{ds}{s-q^2} \frac{s^2
f^2_{Pn(s)}}{(m_1+m_2)^2}\; \frac{dn(s)}{ds}\;.
\label{36}
\end{eqnarray}
Eqs.(\ref{35})-(\ref{36}) give the average correlators for the vector and
pseudoscalar mesons, therefore, due to Eq.(\ref{21}) we state  that
\begin{equation}
\Im m <\Pi^{(hadr)}(q^2)> = \Im m \Pi^{(QCD)}(q^2)\;,
\end{equation}
that gives with account of Eqs.(\ref{27}), (\ref{35}) and
(\ref{36}) at the physical points $s_n =M_n^2$
\begin{equation}
\frac{f_n^2}{M_n} = \frac{\alpha_S}{\pi} \; \frac{dM_n}{dn}
\; \biggl(\frac{4\mu}{M}\biggr)^2\;, \label{38}
\end{equation}
where in the limit of the heavy quarks we use, that for the lightest resonances
one has
\begin{equation}
m_1 +m_2 \approx M\;,
\end{equation}
so that
\begin{equation}
f_{Vn} \simeq f_{Pn} = f_n\;.
\end{equation}
Thus, one can conclude that the QCD sum rules give for the heavy quarkonia
the identity of the $f_P$ and $f_V$ values for  the lightest pseudoscalar
and vector states.

Eq.(\ref{38}) differs from the ordinary sum rule scheme because it does not
contain the parameters, which are external to QCD. The quantity
$dM_n/dn$ is purely phenomenological. It defines the average mass difference
between the nearest levels with the identical quantum numbers.

Note, in the Borel sum rules, the derivative procedure over $\sigma$ gives
the possibility to find both  the constants $f$ and the bound state mass
versus the current quark mass choice.

It must be noted, that the approximation made implies that we neglected
the continuum contribution in the resonant region, and this assumption is valid
for the lightest states only.

The relations, connecting $f$ with the $dM_n/dn$ value, were derived in some
other ways, so in the quasiclassical approximation \cite{11},
in the sum rule analysis with the use of the Euler-McLohren transformation
\cite{12} and by the double action of the Borel transformation \cite{13}.

\section{Sum rule analysis}

As it has been noted in Introduction, the phenomenological properties
of the heavy quark potential lead to the independence of the quarkonium
state density on the  heavy quark flavours (see eq.(\ref{14n})).
Thus, in accordance with eq.({\ref{38}) and in the leading approximation with
no account of the logarithmic and power corrections, one can draw the
conclusion, that for the leptonic constants of the S-wave quarkonia
the scaling relation takes place
\begin{equation}
\frac{f^2}{M}\; \biggl(\frac{M}{4\mu}\biggr)^2 = const.\;, \label{law}
\end{equation}
independently of the heavy quark flavours, so that the constant in the right
hand side of eq.(\ref{law}) is determined by the expression
\begin{equation}
const. = \frac{\alpha_S}{\pi} \; \frac{dM_n}{dn}\;.
\end{equation}
In ref.\cite{10}, for the numerical estimate of the constant we have used
the data on the average mass difference of the S-wave bottomonium
\begin{equation}
<\frac{dM}{dn}> = \frac{1}{2}((M_{\Upsilon'}-M_\Upsilon)+(M_{\Upsilon''}-
M_{\Upsilon'}))\;,
\end{equation}
and the flavour-independent value of $\alpha_S$, determining the Coulomb
term of the Cornel potential,
\begin{equation}
\alpha_S \simeq 0.36\;.
\end{equation}
In the case of the quarkonia with the hidden flavour, one has $4\mu/M=1$.
So, the calculated values of the leptonic constants for the $\Upsilon$-,
$\psi$- and $\phi$-mesons are presented in table \ref{t2}, and they are
in a good agreement with the experimental values of these quantities.

\begin{table}[t]
\caption{The Experimental Values of the Leptonic Constants (in MeV)
for the Quarkonia in Comparison with the Estimates of Present Model.}
\label{t2}
\begin{center}
\begin{tabular}{||l|c|c||}
\hline
quantity & exp. & present \\
\hline
$f_\phi$ & $232\pm5$ & $230\pm25$ \\
$f_\psi$ & $409\pm13$ & $400\pm40$ \\
$f_\Upsilon$ & $714\pm14$ & $700\pm70$ \\
\hline
\end{tabular}
\end{center}
\end{table}

For the heavy $(\bar b c)$  quarkonium with the open charm and beauty,
one can easily find the estimate
\begin{equation}
f_{B_C} = 460\pm 60\;\;MeV\;,\label{bc}
\end{equation}
where the uncertainty is caused by the ambiguity in the choice of the quark
masses \cite{9}
\begin{eqnarray}
m_c & = & 1.4\div 1.8\;\;GeV\;,\\
m_b & = & 4.6\div 5.2\;\;GeV\;.
\end{eqnarray}
In the potential models \cite{14,15,16,17}, the mass estimates for the
S-wave levels of the $B_c$ mesons agree with the expected behaviuor, when
the mass difference is practically the same, as for the families of the
charmonium and the bottomonium.

The $f_{B_C}$ value (\ref{bc}) is in an agreement with the other estimates,
obtained in the framework of both the potential models
\cite{14,15,16,18,19,20,21} and the QCD sum rules \cite{13a,15,17,22}, where,
in the other schemes, a large spread in the $f_{B_C}$ predictions takes
place due to the ambiguity in the choice of the hadronic continuum threshold,
the number of the spectral density momentum or the Borel parameter.

Further, in the limit case of $B$- and $D$-mesons,
when the heavy quark mass is much greater than the light quark mass
$m_Q \gg m_q$, one has
$$
\mu \simeq m_q
$$
and
\begin{equation}
f^2\;M = \frac{16 \alpha_S}{\pi}\;\frac{dM}{dn}\;\mu^2\;.\label{18a}
\end{equation}
Then it is evident that at one and the same $\mu$ one gets
\begin{equation}
f^2\;M = const. \label{19a}
\end{equation}
Scaling law (\ref{19a}) is very well known in EHQT
\cite{1} for mesons with a single heavy quark ($Q\bar q$),
and it follows, for example, from the identity of the $B$- and $D$-meson
wave functions in the limit, when infinitely heavy quark can be considered
as a static source of gluon field.

In our derivation of eqs.(\ref{18a}) and (\ref{19a}) we have neglected
power corrections over the inverse heavy quark mass.
Moreover, we have used the presentation about the light constituent quark
with the mass, equal to
\begin{equation}
m_q \simeq 330\;\;MeV\;, \label{20a}
\end{equation}
so that this quark has to be considered as nonrelativistic one $v \to 0$,
and the following conditions take  place
\begin{equation}
m_Q +m_q \approx M^{(*)}_{(Q\bar q)}\;,\;\;m_q \ll m_Q\;, \label{21a}
\end{equation}
and
\begin{equation}
f_{V} \simeq f_{P} = f\;.
\end{equation}

In agreement with eqs.(\ref{18a}) and (\ref{20a}), one finds the
estimates\footnote
{In ref.\cite{10} the dependence of the S-wave state density
$dn/dM_n$ on the reduced mass of the system with the Martin potential
has been found by the Bohr-Sommerfeld quantization, so that at the step from
($\bar b b$) to ($\bar b q$), the density changes less than about 15\%.}
\begin{eqnarray}
f_{B^{(*)}} = 120 \pm 20\;\;MeV\;, \\
f_{D^{(*)}} = 220 \pm 30\;\;MeV\;,
\end{eqnarray}
that is in an agreement with the estimates in the other schemes of the QCD sum
rules \cite{23}.

Thus, in the limits of $4\mu/M=1$ and $\mu/M \ll 1$, scaling law (\ref{law})
is consistent.

In ref.\cite{13a} the sum rule scheme with the double Borel transform has
been used, so, it allows one to study effects, related to power corrections
from the gluon condensate, corrections due to nonzero quark velocity and
nonzero binding energy of the quarks in the quarkonium.

The numerical effect from the mentioned corrections considers to be
not large (the power corrections are of the order of 10\%),
and the uncertainty, connected to the choice of the quark masses, dominates
in the error of the $f_{B_C}$ value determination.

\section*{Conclusion}

In the present paper, in the framework of the QCD sum rules and in the
leading approximation, we have considered the scaling properties of the
leptonic constants for the S-wave quarkonia with heavy quarks, and
in the specific scheme, allowing one to use the spectroscopic data on
the quarkonium level density, we have gotten the relation
$$
\frac{f^2}{M}\; \biggl(\frac{M}{4\mu}\biggr)^2 = const.\;,
$$
that is in the good agreement with the experimental data on the
leptonic constants for the $\Upsilon$-, $\psi$- and $\phi$-mesons.
It allows one to predict the $f_{B_C}$ value for the $B_c$ meson,
whose search is processed at LEP and FNAL.

The estimates for the leptonic constants of the $B$ and $D$ mesons
due to the scaling relation is in the agreement with the values,
obtained in the framework of the other schemes of the QCD sum rules.

\end{document}